\documentclass{icrc}

\usepackage{times}
\usepackage{graphicx} 

\begin{document}

\title{A Possible High Altitude High Energy Gamma Ray Observatory in India}
\author[1,2]{R. Cowsik}
\affil[1]{Indian Institute of Astrophysics, Koramangala, Sarjapur Road,
Bangalore 560 034, India}
\author[2]{P. N. Bhat}
\author[2]{V. R. Chitnis}
\author[2]{B. S. Acharya}
\author[2]{P. R. Vishwanath}
\affil[2]{Tata Institute of Fundamental Research, Homi Bhabha Road,
Mumbai 400 005, India}

\correspondence{R. Cowsik (cowsik@iiap.ernet.in)}

\runninghead{Cowsik et al.: Possible high altitude gamma ray observatory}
\firstpage{1}
\pubyear{2001}


\maketitle

\begin{abstract}
        Recently an Indian Astronomical Observatory  has been set up at
Hanle (32$^\circ$ 46$^\prime$ 46$^{\prime\prime}$ N, 78$^\circ$ 57$^\prime$
51$^{\prime\prime}$ E, 4515m amsl)
situated in the high
altitude cold desert in the Himalayas. The Observatory has 2-m aperture
optical-infrared  telescope, recently built by the Indian Institute of
Astrophysics.

        We have carried out systematic simulations for this observation
level  to study the nature of \v Cerenkov light pool generated by gamma ray
and proton primaries incident vertically at the top of the atmosphere. The
differences in the shape of the lateral distributions of \v Cerenkov light with
respect to that at lower altitudes is striking. This arises primarily due
to the proximity of the shower maximum to the observation site. The
limited lateral spread of the \v Cerenkov light pool and near 90\% atmospheric
transmission at this high altitude location makes it an ideal site for
a  gamma ray observatory. This results in a decrease in the gamma ray
energy threshold by a factor of 2.9 compared to that at sea-level.
Several parameters based on density and timing 
information of \v Cerenkov photons, including local and medium range 
photon density fluctuations as well as photon arrival time jitter could be
efficiently used to discriminate gamma rays from more abundant cosmic rays at
tens of GeV energies.
\end{abstract}

\section{Introduction}

Atmospheric \v Cerenkov technique is a well established technique for
study of VHE gamma ray emission from astronomical sources. This technique
has been successfully exploited by several experiments such as Whipple,
CAT, CANGAROO, HEGRA, TACTIC $etc$ based on imaging technique as
well as by CELESTE, STACEE, SOLAR-2, GRAAL, PACT $etc$ based on wavefront 
sampling technique (Ong, 1998).
Next generation experiments including large imaging telescope, like MAGIC, as
well as arrays of imaging telescopes such as VERITAS and HESS are under
construction. These experiments as well as wavefront sampling experiments
with large collection area such as CELESTE, SATCEE $etc$ are expected to
achieve low energy threshold of the order of few tens of $GeV$. Alternatively,
it is possible to decrease energy threshold by conducting an experiment at
higher observation altitudes. All the existing experiments are being carried 
out at
altitudes of upto 2.5 $km$. Here we investigate the feasibility of an
experiment based on wavefront sampling technique at a location called
Hanle situated in the cold desert in the Himalayas at an altitude of about
4.5 $km$, based on simulation studies.
 
Mt. Saraswati in Hanle is an exceptionally fine astronomical site offering 
about 260 spectroscopic nights per year, with uniform coverage of all right
ascensions, low precipitable water vapour ($\sim 1~mm~cm^{-2}$), low aerosol
content and extinction $(\sim 0.1^m ~ in~V~band)$, low sky brightness $21^m.5(V)
~ arcsec^{-2}$ and median seeing $< 1''$. Moreover it is situated right in the 
middle of the gap between Woomera ($137^\circ~ E$) in Australia and La Palma 
($20^\circ~ W$).

\section{Lateral distributions of \v Cerenkov photons}

We have simulated a large number of air showers generated by $\gamma-$rays 
and protons of various primary energies using CORSIKA (Heck et al., 1998).
The \v Cerenkov radiation produced by the secondary charged particles
in the shower within the bandwidth of 300-650 $nm$ is propagated to the
observation level. 
Location and altitude appropriate for Hanle 
are used in simulations. An array of 357 telescopes, each
consisting of seven mirrors with a total area of 4.45 $m^2$ per telescope, 
spread over an area of 400 m $\times$ 400 m is considered. 
All the showers are vertically incident at the top of the atmosphere,
with shower core chosen to be at the centre of the array. Typically
100 showers were simulated for higher energy $\gamma-$rays (50 and
500 GeV) and protons of energies 150 GeV and 1 TeV. For lower
energy primaries, i.e., $\gamma-$rays of energy 1 and 10 GeV and
for protons of energy 15 and 50 GeV, 500 showers were simulated.
Energies of primaries are chosen so that $\gamma-$ray and proton showers
have comparable \v Cerenkov yields.

\begin{figure}[t]
\vspace*{2.0mm} 
\includegraphics[width=8.3cm]{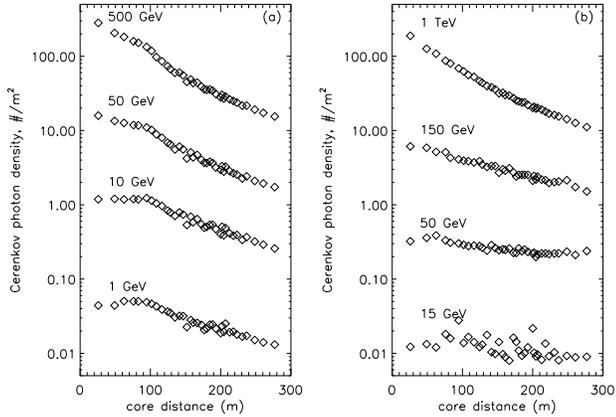}
\caption{Average \v Cerenkov photon density at Hanle as a function of 
core distance for showers initiated by (a) $\gamma-$rays of energies 1, 10,
50 and 500 GeV and (b) protons of energies 15, 50, 150 GeV and 1 TeV. 
Distributions are averaged over 500 showers for lower energy primaries
and over 100 showers for higher energy ones.}
\end{figure}

Figure 1 shows the average \v Cerenkov photon density as a function
of core distance for showers initiated by $\gamma-$rays and protons
of various energies. Lateral distributions from $\gamma-$ray primaries
indicate presence of proverbial hump at a core distance of about 90 m,
due to effective focusing of \v Cerenkov photons from a range of
altitudes. However, this hump is somewhat less prominent compared to
that seen at lower altitudes, for example, the one seen at observation
altitude of 1 km (Chitnis and Bhat, 1998). Also the density 
distribution within hump is not as flat as in the case of lower observation
altitudes. Dilution of the hump at higher primary energies as well as
at higher altitudes is an expected feature (Rao and Sinha, 1988).
Also the comparison of lateral distributions show that the \v Cerenkov 
photon density near the shower core at Hanle is higher by a factor of about 
5-6 compared to that at sea-level, for a given primary energy. 
Wavelength dependent atmospheric attenuation of \v Cerenkov photons is 
not taken into consideration here. This higher photon density as well as 
the smaller distance to hump from shower axis at Hanle is due to the 
compactness of shower at higher altitudes. This will reduce the energy 
threshold of the experiment appreciably compared to same array at lower 
altitudes.

\section{Gamma-hadron separation}

All atmospheric \v Cerenkov experiments have to deal with a substantial
background from air showers generated by cosmic rays emulating those
initiated by $\gamma-$ray primaries. It is necessary to incorporate the
methods for effective rejection of this background for improving signal
to noise ratio. In imaging experiments background rejection is based on
differences in shapes and orientations of images produced by these two
species (Fegan, 1997). Whereas in experiments based on wavefront sampling 
technique
parameters based on arrival time of \v Cerenkov shower front and 
\v Cerenkov photon density at various locations in \v Cerenkov pool can 
be used for discrimination. The usefulness
of these techniques at lower observation altitudes has already demonstrated 
(Chitnis and
Bhat, 2001; Bhat and Chitnis, 2001). Here we study the effectiveness
of these parameters at Hanle altitude.

We use quality factor as a figure of merit to distinguish between
$\gamma-$ray and proton initiated showers. If is defined as
\begin{equation}
Q_f={{N_a^{\gamma}} \over {N_T^{\gamma}}} \left( {{N_a^{pr}} \over {N_T^{pr}}} \right) ^{-{{1} \over {2}}} 
\end{equation}
where $N_a^{\gamma}$ is the number of $\gamma-$rays accepted (i.e. below
threshold), $N_T^{\gamma}$ is the total number of $\gamma-$rays,
$N_a^{pr}$ is the number of protons accepted and $N_T^{pr}$ is the total
number of protons.

\subsection{GHS based on timing information}

We have examined the applicability of three types of parameters based on
\v Cerenkov photon arrival times at various
locations in \v Cerenkov pool, {\it viz}, 1. the curvature of shower
front, 2. shape of \v Cerenkov pulse at the telescopes and 3. relative
arrival time jitter. For details of these parameters see Chitnis and
Bhat (2001). For a given shower, mean arrival times of shower front
at various core locations are fitted with a spherical front.
Radius of curvature of this shower front is found to be roughly equal to the 
height
of the shower maximum from the observation level. It also
provides moderate discrimination against cosmic ray showers. This is
mainly because of the interaction length of hadrons in the atmosphere
being around twice the radiation length. Hence the hadron initiated
showers reach shower maximum deeper in the atmosphere compared to a 
$\gamma-$ray initiated showers.  
Optimum quality factor, derived using radius of curvature as a parameter
is given in Table 1 which is self explanatory.  Distributions 
of fitted radii for showers initiated by 500 GeV $\gamma-$rays and 
1 TeV protons are shown in Figure 2. Threshold radius for optimum quality
factor is indicated by dashed line.

Second parameter investigated is related to pulse shape. 
As in the case of lower observation altitudes, the 
\v Cerenkov pulse decay time gives reasonably good discrimination, whereas rise 
time and pulse width are not much effective. Discrimination is best in the
vicinity of hump region at all altitudes. Quality of discrimination is
somewhat inferior at Hanle altitude compared to lower altitudes due to the 
dilution of hump at this observation level. The \v Cerenkov photon density
for lower energy $\gamma-$ray primaries remains almost constant until a
core distance of $\sim$ 110 m. Hence quality factors have been calculated 
using telescopes within a  core distance of 110 m.
Quality factor using decay time of \v Cerenkov pulse is listed in Table 1. 
Based on decay time alone it is possible to reject
about 76\% of proton showers, retaining about 68\% of $\gamma-$ray showers.
Distributions of decay times for both the primaries are shown in Figure 2
along-with the threshold corresponding to optimum quality factor.

Third parameter is the relative timing jitter. This is the ratio of 
RMS of average arrival times of \v Cerenkov photons at seven mirrors of 
the telescope to the mean of seven averages. Due to the differences in 
kinematics, cosmic ray showers are expected to have higher timing jitter 
compared 
to $\gamma-$ray showers. Also relative jitter is found to be roughly 
independent of core distance. Quality factor based on timing jitter for 
the showers generated by 500 GeV $\gamma-$rays and 1 TeV protons for 
telescopes within a core distance of 110 m are listed 
in Table 1. Distributions of relative timing jitter for both the species
are shown in Figure 2. Threshold value of jitter for optimum quality
factor is also indicated. 
Based on the arrival time jitter, it is possible to reject about 
97\% of proton showers retaining about 49\% of $\gamma-$ray showers.

\begin{figure}[t]
\vspace*{2.0mm} 
\includegraphics[width=8.3cm]{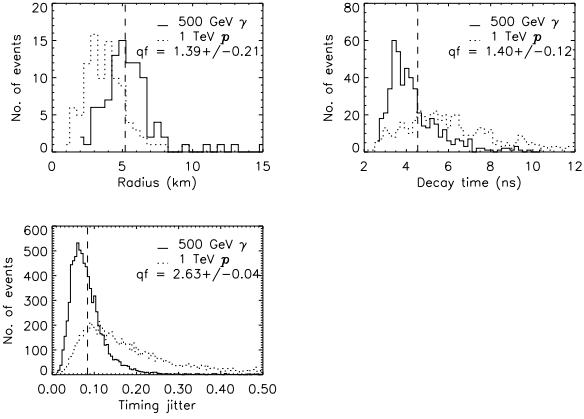}
\caption{
Distributions of three primary species sensitive parameters based on \v Cerenkov
photon timing information $viz.$ radius of curvature of the shower front, pulse
decay time and relative arrival time jitter. The derived quality factors and the
primary energies considered here are indicated.
}
\end{figure}

\begin{table}
\caption{Gamma-hadron separation for showers initiated by 500 GeV 
$\gamma-$rays and 1 TeV protons at Hanle}
\begin{tabular}{lllll}
\hline
Parameter &  Threshold   & Quality factor & Fraction  & Fraction  \\
          &  value       &                & of & of \\
          &              &                & accepted    & accepted \\
          &              &                & $\gamma-$rays & protons \\
\hline
Shower front & 5.2 km & 1.39 $\pm$ 0.21 & 0.577 & 0.173 \\
curvature    &&&&\\
Decay time    & 4.54 ns & 1.40 $\pm$ 0.12 & 0.682 & 0.236 \\
of pulse &&&&\\ 
Timing &  0.084 & 2.63 $\pm$ 0.02 & 0.487 & 0.034 \\
jitter &&&&\\
Decay time & 4.54 ns, & 2.25 $\pm$ 0.05 & 0.349 & 0.024 \\
and jitter & 0.084 &&&\\
LDF  &  0.127 & 1.53 $\pm$ 0.03 & 0.803 & 0.276 \\
MDF  &  0.164 & 1.24 $\pm$ 0.09 & 0.386 & 0.097 \\
Flatness & 34.8 & 1.01 $\pm$ 0.05 & 0.963 & 0.902 \\
parameter &&&&\\
LDF and  & 0.127,  & 1.60 $\pm$ 0.15 & 0.338 & 0.045\\
MDF  &   0.164 &&&\\ 
\hline
\end{tabular}
\end{table}

\subsection{GHS based on \v Cerenkov photon density}

There are certain kinematical differences in air showers initiated by
cosmic rays and $\gamma-$rays. These differences originate from those
in first interaction of primary, presence of hadronic secondaries and
muons in cosmic ray showers. As a result, cosmic ray showers are 
expected to show larger density fluctuations compared to $\gamma-$ray 
showers. We have parameterized density fluctuations 
and examined their efficacy for gamma hadron separation. Three types
of parameters have been already studied for lower observation altitudes 
(Bhat and Chitnis, 2001). First parameter considered 
is the local density fluctuations (LDF) or density jitter. Each 
telescope consists of seven mirrors and LDF is the ratio of RMS of
\v Cerenkov photon densities at these mirrors to the mean density. As
in the case of lower altitudes, we find that for Hanle altitude also
LDF is larger for proton showers compared to $\gamma-$ray showers at
all the core distances. Quality factor 
based on LDF for core distance within 110 $m$ is given in
Table 1. 
Based on LDF it is possible to reject about 72\% of proton
showers retaining about 80\% of $\gamma-$ray showers. 

Secondly, we consider medium range density fluctuations (MDF).
As in the case of PACT or Pachmarhi Array of \v Cerenkov telescopes
(Bhat et al., 2001), we assume that the proposed array to be divided 
into the sectors
consisting of six telescopes in each sector. Then MDF is defined as
the ratio of RMS of photon densities recorded at six telescopes to the mean
density. As in the case of lower altitudes, MDF is larger
for proton showers at all core distances. Quality factor 
based on MDF  for core distances within 110 $m$
is listed in Table 1. It is possible to reject about 90\%
of proton showers retaining about 39\% of $\gamma-$ray showers using 
MDF as discriminating parameter.

Third parameter investigated is the well-known flatness parameter. 
Lateral 
distributions from $\gamma-$ray showers are roughly flat within hump
region. Also the lateral distributions from $\gamma-$ray showers are smooth
compared to proton showers. These differences in lateral distributions 
can be parameterized using flatness parameter $\alpha$, which is defined as
\begin{equation}
\alpha={{1} \over {N}} \left[\sum_{i=1}^{N} {{ \left( \rho_i - \rho_0 \right) ^2} \over {\rho_0}} \right]  
\end{equation}
where $N$ : no. of telescopes triggered in sector, $\rho_i$ : photon density 
measured by individual telescopes of a sector and $\rho_0$ : average density.

Lateral distributions from $\gamma-$ray showers are expected to have a 
smaller value of $\alpha$ compared to that from proton generated showers, 
on the average. At lower altitudes we have seen that proton showers 
have larger value of flatness parameter compared to $\gamma-$ray showers
at all distances away from hump (Bhat and Chitnis, 2001). Hence 
flatness parameter serves as a good discriminant at core distances away
from hump. However, at Hanle altitude flatness parameter is not a useful
discriminant as reflected in smaller quality factor listed in Table 1.
This is primarily due to the 
reduction in differences between the lateral distributions of \v Cerenkov
photons generated by the two
species at higher altitudes. LDF and MDF, on the other hand, provide comparable
background rejection at all the observation altitudes.

\section{Discussion and conclusions}
\subsection{\v Cerenkov photon lateral distribution}
It is generally said that the lateral distributions of \v Cerenkov radiation 
from $\gamma $-ray and proton generated showers are distinctly different in
the sense that in the former case it is flat up to about $\sim 140~m$ 
at sea level and
characterized by a hump at that distance while in the latter case it is steeper
and smoother with practically no hump (Rao \& Sinha, 1988). However the 
situation 
changes as the observation altitude increases, since the shower maximum for a
given primary energy comes closer to the observation level. This situation is
similar to the case of increasing primary energy at a given observation
altitude. Thus the prominence of hump decreases with increasing altitude.
For the same reason the core distance at which the hump appears also decreases
with increasing observation level. At an observation altitude of 4500 $m$, 
where the grammage is $\sim 598~g cm^{-2}$, the radius of curvature (or the
height $\gamma $-ray shower maximum from observation level) 
is around $5~km$. The \v Cerenkov angle at shower
maximum is around $1^\circ$ and the expected position of hump is $\sim 90~m$
purely from geometric considerations which agrees well with figure 1. The 
proximity of the shower maximum to the observation level becomes more sever
for higher energy $\gamma $-ray primaries and the hump almost disappears.
Here the contribution from higher energy electrons closer to the observation
level becomes appreciable at near core distances because of which at higher
observation altitudes the hump is seen in the case of lower energy
primaries only.

Another feature of the \v Cerenkov photon lateral distributions is that they 
become increasingly flatter with decreasing primary energy. The flattening is
far more significant for proton primaries as compared to $\gamma $-ray 
primaries. As a result the pool size increases with lowering primary energy
which is a consequence of significantly larger number of photons arriving
at larger angles. When the lateral distribution curves are generated with a
finite focal point mask, the density as well as the total 
number of photons detected reduces significantly for proton primaries. For
example, the fractions of photons detected when a $5^\circ$ mask is in use are
64.3\% and 33.2\% respectively for 50 GeV \& 15 GeV protons. Similar fractions
for $\gamma $-ray primaries are  90.1\% and 96.4\% respectively for 10 \& 1 GeV 
$\gamma $-rays. As a result, at lower primary energies, the use of a focal 
point mask provides a simple discrimination against hadrons.

In addition, the atmospheric attenuation of \v Cerenkov photons at Hanle
altitude is $\sim 14$\% as compared to $\sim 50$\% at sea-level. The ratio  of 
\v Cerenkov yield  for high energy $\gamma $-rays to that of protons of same
energy increases exponentially with decreasing energy (Ong, 1998). Combined 
with
increased photon density due to reduced lateral spread of the pool makes a 
high altitude observatory like Hanle an ideal site for GeV $\gamma $-ray 
astronomy.
The above two considerations are expected to reduce the $\gamma-$ray energy
threshold by a factor of $\sim$ 2.9 compared to that at sea level.

\subsection{Gamma - hadron separation}

Because of the proximity of the shower maximum at higher observation altitudes,
 radius of curvature is more
sensitive to primary species as compared to lower observation levels. However
certain parameters like the pulse decay time, which is more sensitive to
the presence of hump, is relatively less sensitive to the primary species 
compared to that at lower observation altitudes. Third parameter, $viz.$ the
relative timing jitter is comparable to that at lower levels. Combining the
second and the third parameter in tandem makes $\gamma $-hadron separation more 
efficient at higher observation altitude.
As can be seen from table 1, using these parameters in tandem it is possible to
reject about 98\% of proton showers retaining about 35\% of $\gamma-$ray
showers.

Similarly, among the density based parameters, $\alpha$ is less sensitive 
at higher energies because of the similarity between the lateral distributions
of $\gamma $-rays and protons. However efficiencies of LDF and MDF as 
discriminants are not very sensitive to the observation altitudes.  
Background rejection can be improved further by applying various parameters
in tandem. MDF and flatness parameter are very similar in definition,
both calculated using \v Cerenkov photons densities at each telescope in the
sector. Hence these parameters are not strictly independent. LDF, on the
other hand, is density jitter in the telescope itself and hence independent
of MDF or flatness parameter.  As can be seen from the table 1, if one uses 
these 
two parameters in tandem then it is possible to reject about 95\% of proton
showers retaining about 34\% of $\gamma-$ray showers. 

By exploiting the advantages of the high observation altitude and low energy 
characteristics of \v Cerenkov emission in the atmosphere, it is possible to 
achieve very low energy threshold as well as an excellent gamma-hadron 
discrimination without using bulky and expensive hardware.


\end{document}